\documentclass[reprint,amsmath,amssymb,superscriptaddress,aps,prl]{revtex4-2}

\usepackage{graphicx}
\usepackage{dcolumn}
\usepackage{bm}
\usepackage[usenames,dvipsnames]{color}
\usepackage{epstopdf}
\usepackage{lipsum}
\newcommand\Wen{\mbox{\textit{We}}}
\usepackage{hyperref}

\newcommand{\revP}[1]{\textcolor{black}{#1}}

\begin{document}

\title{The Leidenfrost effect as a directed percolation phase transition}
\author{Pierre Chantelot}
\email{p.r.a.chantelot@utwente.nl}
\affiliation{Physics of Fluids Group, Max Planck Center Twente for Complex Fluid Dynamics, MESA+ Institute, and J. M. Burgers Center for Fluid Dynamics, University of Twente, P.O. Box 217, 7500AE Enschede, Netherlands}
\author{Detlef Lohse}
\email{d.lohse@utwente.nl}
\affiliation{Physics of Fluids Group, Max Planck Center Twente for Complex Fluid Dynamics, MESA+ Institute, and J. M. Burgers Center for Fluid Dynamics, University of Twente, P.O. Box 217, 7500AE Enschede, Netherlands}
\affiliation{Max Planck Institute for Dynamics and Self-Organisation, Am Fassberg 17, 37077 Göttingen, Germany}

\begin{abstract}

  Volatile drops deposited on a hot solid can levitate on a cushion of their own vapor, without contacting the surface.
  We propose to understand the onset of this so-called Leidenfrost effect through an analogy to non-equilibrium systems exhibiting a directed percolation phase transition.
  When performing impacts on superheated solids, we observe a regime of spatiotemporal intermittency in which localized wet patches coexist with dry regions on the substrate.
  We report a critical surface temperature, which marks the upper bound of a large range of temperatures in which levitation and contact coexist.
  In this range, with decreasing temperature, the equilibrium wet fraction increases continuously from zero to one.
  Also, the statistical properties of the spatio-temporally intermittent regime are in agreement with that of the directed percolation universality class.
  This analogy allows us to redefine the Leidenfrost temperature and shed light on the physical mechanisms governing the transition to the Leidenfrost state.

\end{abstract}

\maketitle

Johann Gottlob Leidenfrost reported, in his 1756 treatise, that volatile liquids deposited on superheated substrates can float on a cushion of their own vapor \cite{leidenfrost1756}. 
Levitation occurs when the substrate is heated above the Leidenfrost temperature $T_L$.
In this non-contact situation, the heat transfer is dramatically reduced, with often dramatic consequences in cooling applications \cite{moreira2010,hamdan2015,breitenbach2018} or diesel engines \cite{law1982,cho1985}, and the drop lifetime is consequently enhanced \cite{biance2003}.
The Leidenfrost temperature is classically defined as the temperature which maximises the lifetime of a drop deposited on a hot substrate \cite{gottfried1966,biance2003}.
Experiments show that $T_L$ depends on the thermal properties of the liquid and the solid \cite{gottfried1966,baumeister1973,bernardin1999}, surface roughness \cite{bernardin1999,kim2011}, wettability \cite{vakarelski2012,bourrianne2019}, and liquid impurities \cite{bernardin1999,cui2003}.
The measured values of $T_L$ are typically $100^\circ$C larger than the liquid's boiling point $T_b$, and vary greatly. For example, the Leidenfrost point of distilled water on a polished metal plate has been reported to range from $150^\circ$C \cite{biance2003} to $300^\circ$C \cite{gottfried1966}.

\begin{figure}[b]
  \centering
   \includegraphics[width=0.475\textwidth]{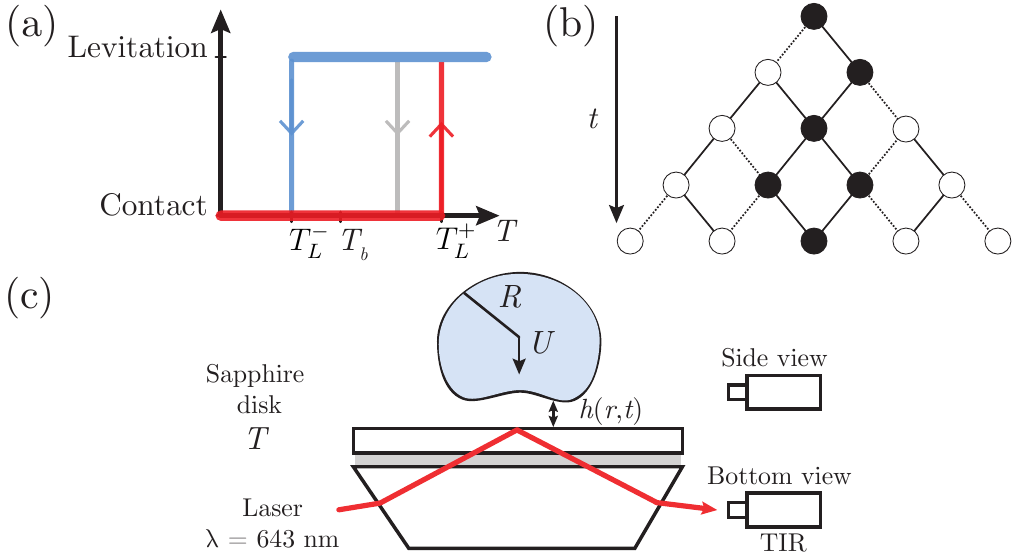}
   \caption{\label{fig1}
   (a) Schematic phase diagram for levitation and contact. Both states can be (locally) observed for substrate temperatures between $T_L^-$ and $T_L^+$ where they coexist.
   (b) Sketch of the dynamics of directed bond percolation.
   (c) Ethanol drops with equilibrium radius $R$ and velocity $U$ impact a sapphire substrate with temperature $T$. We record side views and use total internal reflection (TIR) imaging to measure the thickness $h(r,t)$ of the gas film squeezed between the liquid and the solid (sketch not to scale).
  }
\end{figure}

In their seminal work, Baumeister \emph{et al.} \cite{baumeister1966} rationalized the uncertainty in the determination of $T_L$ by noticing that initially floating water drops can survive when the substrate is cooled down to the liquid's boiling point.
This observation shows that the vapor layer can support the liquid even for vanishing superheat as predicted by lubrication models \cite{sobac2014}.
They also reported that perturbations can terminate levitation for temperatures ranging from $100^\circ$C to $200^\circ$C, the later value being consistent with a typical value of $T_L$, suggesting that the levitated drops are metastable for a large temperature range whose bounds we denote $T_L^-$ and $T_L^+$ (figure \ref{fig1}a).
Further experiments \cite{kolinski2014_2,harvey2020,vanlimbeek2021,chantelot2021} confirmed the metastability of the levitated state and showed that the lower threshold of the Leidenfrost temperature $T_L^-$ is controlled by hydrodynamic instabilities of the gas layer, or by surface contamination and/or roughness.
On the contrary, \revP{measuring $T_L^+$ is still challenging because of the hysteresis introduced by the hydrodynamic stability of the vapor layer and only recently attempts have been made to derive $T_L^+$ from a stability analysis at the nanoscale \cite{Zhao2020}.}
The physical mechanisms that set $T_L^+$ thus remain elusive.

In this Letter, we propose an analogy between the transition to the Leidenfrost state and non-equilibrium systems exhibiting a phase transition into an absorbing state in the directed percolation (DP) universality class.
Such transitions arise from simple mechanisms, and have thus been used to model a wide range of non-equilibrium phenomena from the spread of epidemics, to catalytic reactions, and the onset of turbulence \cite{hinrichsen1999,takeuchi2007,henkel2008,barkley2016,kerswell2018}.
In figure \ref{fig1}b, we sketch the dynamics of a prototypical realisation of DP, bond percolation and illustrate them with the example of the onset of turbulence in subcritical shear flows, for which a similar analogy was formulated by Pomeau \cite{pomeau1986} and later verified by experiments and simulations \cite{lemoult2016,sano2016,chantry2017,avila2011}.
An active state (in black) spreads to its nearest neighbours through connected bonds with a probability $P$.
Active sites (\emph{i.e}, turbulent patches) cannot spontaneously arise, but only occur through interactions with already active domains. This condition reflects the absorbing nature of the inactive state and, in our example, stems from the linear stability of the flow.
The evolution of the system is controlled by a single parameter, the probability $P$, whose equivalent is the Reynolds number \revP{for the onset of turbulence in subcritical shear flows}.
There is a critical probability below which the system returns to a fully inactive (\emph{i.e}, laminar) state and above which a second order phase transition occurs and the fraction of active sites (\emph{i.e}, the turbulent fraction) is always greater than zero.
\revP{Non-equilibrium wetting models provide insight on how to map a DP process to the Leidenfrost effect by suggesting to understand the onset of levitation, at $T_L^+$, as an unpinning transition.
Indeed, such models argue that, in the presence of a short-range attractive force, a second order phase transition in the DP universality class marks the boundary of a regime where contact (active) and levitated (inactive) patches coexist on the substrate.
This boundary corresponds to the detachment of the liquid/gas interface \cite{hinrichsen2003,ginelli2003}, in agreement with the observed metastability of the levitated state, which reflects its absorbing nature.}

We now investigate whether this analogy holds, using drop impact experiments as a method to perturb the metastable levitated state, and thereby directly observe the behavior of contact points using Total Internal Reflection (TIR) imaging \cite{kolinski2014,shirota2017}.
We first evidence a \emph{continuous} transition towards the Leidenfrost state when varying the substrate temperature $T$ -- not a sharp one, as often assumed -- and then discuss whether this transition lies in the directed percolation universality class.

Our experiments, sketched in figure \ref{fig1}c, consist in impacting ethanol (density $\rho = 789$ kg/m$^3$, and surface tension $\gamma = 22$ mN/m) drops on an optically smooth sapphire substrate. We choose this liquid-solid combination as the excellent thermal conductivity of sapphire ($k = 35$ W/K/m) approximates isothermal substrate conditions and allows to neglect vapor cooling \cite{vanlimbeek2016,vanlimbeek2017}.
The substrate temperature is set to a fixed superheat $\Delta T = T - T_b$, where $T_b = 351$ K is the boiling point of ethanol, ranging from $69 \pm 2$ K to $104 \pm 2$ K, using a proportional-integral-derivative controller, and calibrated using a surface probe.
Drops with radius $R = 1.1$ mm are released from a calibrated needle, whose height is adjusted to obtain the desired impact velocity $U$.
\revP{In this paper, we harness impacts to generate quasi-two-dimensional vapor films (with thickness $h \ll R$) \footnote{We expect the gas film squeezed under the drop to be mostly composed of vapor after a time on the order of 0.5 ms \cite{chantelot2021}, although we cannot exclude air remains trapped in this contact situation.}, and to trigger liquid/solid contact through surface contamination.
We will not discuss the influence of liquid inertia and the so-called dynamic Leidenfrost effect \cite{tran2012}, where impacting drops have large impact velocities: we employ low impact velocities $U$ between $0.6$ m/s and $0.8$ m/s (\emph{i.e} $We = \rho R U^2/\gamma = 14 - 25$), close to the transition from levitation to contact.}
We record side views at 20 kHz, from which we obtain the impact velocity $U$ and the drop radius $R$.
Simultaneously, we observe the occurrence of contact using a second synchronized bottom-view high-speed camera, connected to a long distance microscope (with $15$ $\mu$m/px resolution), which records TIR snapshots at 200 kHz. The grayscale intensity of these snapshots yields quantitative thickness measurements which allows us to distinguish wet from dry areas (see Supplemental Material for details of the optical setup and image processing \footnote{See Supplemental material at [url], which includes Refs. \cite{richard2000_2,bottin1998}}).
\begin{figure}
  \centering
   \includegraphics[width=0.475\textwidth]{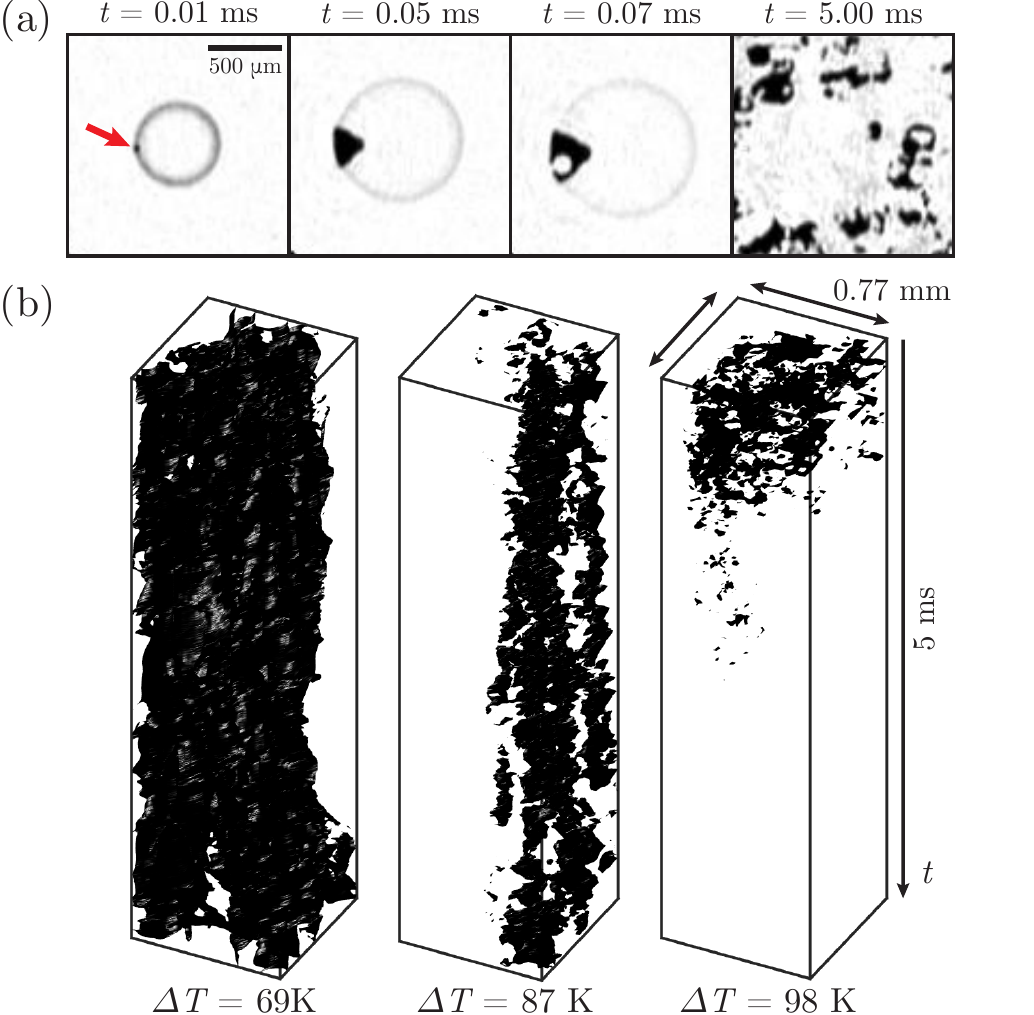}
   \caption{\label{fig2}
   (a) TIR snapshots for the impact of a drop with radius $R = 1.1$ mm and impact velocity $U = 0.65$ m/s (\emph{i.e}, $\Wen = 17$) for an imposed superheat $\Delta T = 87$ K. Contact nucleates at $t = 0.01$ ms (red arrow) and spreads until it starts receding ($t=0.07$ ms).
   (b) Spatiotemporal binarized representations of the wet regions for three superheat below ($\Delta T = 69$ K), close to ($\Delta T = 87$ K), and above ($\Delta T = 98$ K) the critical point. We show a square of 0.77 mm $\times$ 0.77 mm within the observation area during 5 ms (time goes downwards).}
 \end{figure}

In figure \ref{fig2}a, we show TIR snapshots recorded during an impact ($R = 1.1$ mm, $U = 0.65$ m/s) on a substrate heated 87 K above the boiling point of ethanol. Liquid/solid contact nucleates at a single point (red arrow, $t = 0.01$ ms) and the liquid spreads. Yet, at $t=0.07$ ms, the liquid front recedes, in contrast to observations at ambient temperature where dewetting does not occur \cite{kolinski2012,kolinski2019}.
The subsequent spreading and receding dynamics lead to the creation of an extended two-dimensional wetting pattern ($t = 5.00$ ms), called transition boiling pattern \cite{shirota2016}.
This observation highlights the existence of the mechanisms necessary for the proposed analogy: active wet patches evolve by spreading on initially inactive dry regions, and, in the presence of superheat, also recede to the inactive state showing the characteristics of a regime of spatiotemporal intermittency (see Movie S1), similarly as at the transition from the laminar to the turbulent regime in pipe flow \cite{barkley2016}.
In our situation, the order parameter is the wet fraction, $\rho(t)$, that we measure by binarising TIR snapshots using a quantitative thickness criterion, $h(r,t) < h_c = 66$ nm, which allows us to discriminate the wet from the dry regions.
\revP{Varying $h_c$ in the range 40 nm to 100 nm (implying a variation of the intensity threshold by a factor 4 \cite{shirota2017}) does not affect the scaling properties.}
We first qualitatively study the influence of $\Delta T$ by creating binarised spatiotemporal diagrams showing a 0.77 mm $\times$ 0.77 mm area inside the wetted region during 5 ms for three different superheat (figure \ref{fig2}b).
For $\Delta T = 69$ K, wet patches, initially coexisting with dry areas, invade a large fraction of the domain.
In contrast, they eventually disappear for $\Delta T = 98$ K, revealing the existence of an unpinning transition of the liquid/gas interface.
Close to the critical point (later determined to be $\Delta T_c = 87.7 \pm 1.3$ K) at $\Delta T = 87$ K, wet regions persist among large dry areas, suggesting a continuous phase transition.

\begin{figure}
  \centering
   \includegraphics[width=0.475\textwidth]{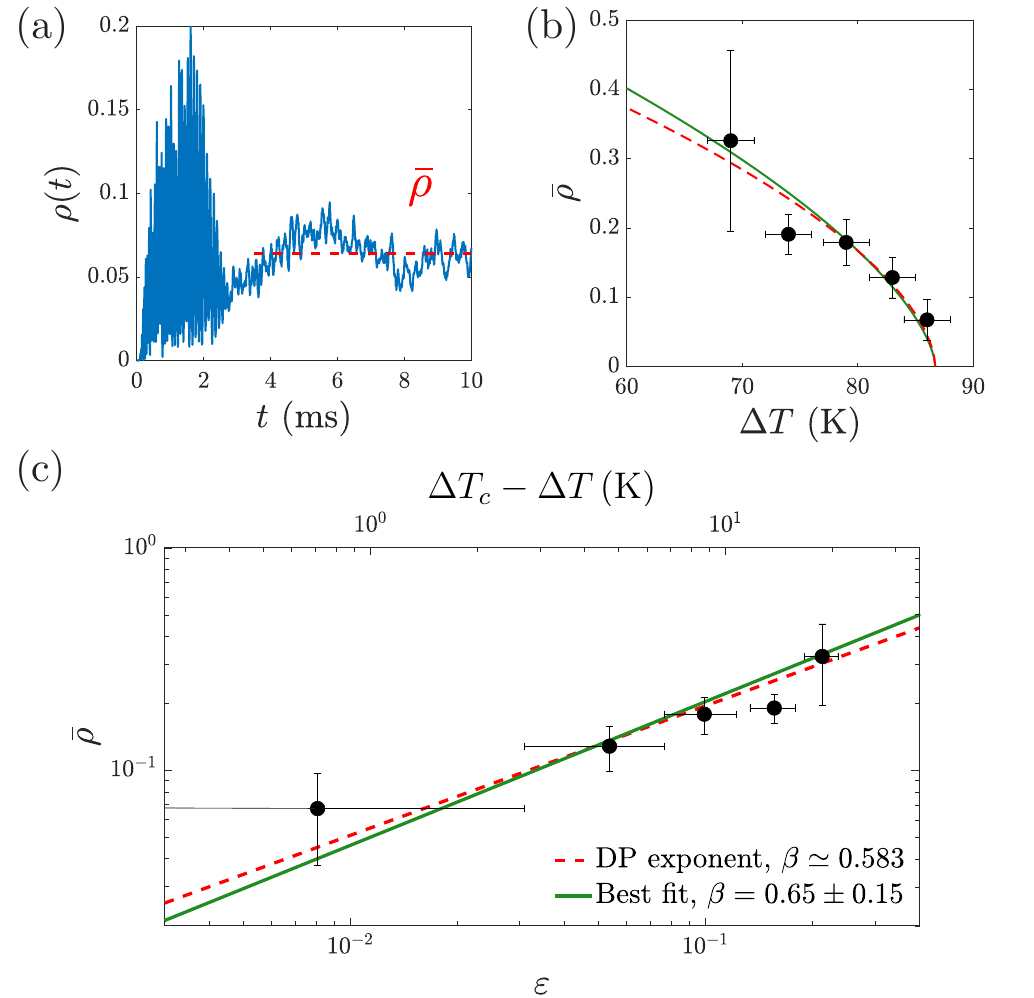}
   \caption{\label{fig3}
   (a) Temporal evolution of the wet fraction $\rho(t)$ during an impact with $R = 1.1$ mm, $U = 0.65$ m/s (\emph{i.e} \Wen = 17), and $\Delta T = 87$ K. After an initial transient, the wet fraction reaches a plateau value, $\bar{\rho}$.
   (b) Mean wet fraction $\bar{\rho}$ in the steady state as a function of the imposed superheat $\Delta T$. The vertical and horizontal error bars represent the standard deviation of $\rho(t)$ and the accuracy of the thermocouple, respectively.
   (c) Same data plotted as a function of $\varepsilon = (\Delta T_c - \Delta T)/\Delta T_c$ and on a log-log scale. The lines are fits of the form $\bar{\rho} \sim \varepsilon^\beta$ where $\beta$ is the DP exponent in (2+1) dimensions (dashed red line) and the best fit exponent (solid green line).}
\end{figure}

We now quantitatively investigate the variation of the wet fraction when varying the superheat $\Delta T$ from 69 K to 87 K.
We fix the impact velocity $U = 0.65$ m/s, and choose a $105$ px $\times$ $105$ px observation area centered on the impact location which size is limited by the extent of the wetted region under the drop.
In figure \ref{fig3}a, we show the temporal evolution of the wet fraction $\rho(t)$ during an impact for $\Delta T = 87$ K.
After an initial transient, during which we observe large oscillations, the wet fraction reaches a plateau, whose value we denote $\bar{\rho}$, and that lasts for at least $7$ ms.
We interpret this plateau as a steady state spatiotemporal intermittency regime, although its observation time is limited in our experiments by (i) the drop dynamics that reduce the wetted area and (ii) substrate cooling
(see Supplemental Material for \revP{a discussion of the steady state approximation and of these two effects}).
In figure \ref{fig3}b, we plot the steady state order parameter $\bar{\rho}$, averaged over ten impacts for each $\Delta T$, as a function of the superheat.
The wet fraction continuously decreases to zero with increasing superheat, evidencing a second order phase transition characterised by a power law scaling $\bar{\rho} \sim \varepsilon^\beta$, where $\varepsilon = (\Delta T_c - \Delta T)/\Delta T_c$, and $\Delta T_c$ is the critical point (figure \ref{fig3}c).
We first determine $\Delta T_c$ by linearly fitting the logarithms of $\bar{\rho}$ and $\varepsilon$ assuming $\beta = 0.583$, which is the value predicted for DP in (2+1) dimensions \cite{grassberger1996}.
This fit (dashed red line) yields the estimate $\Delta T_c = 87.7 \pm 1.3$ K, that we use as an input to obtain the fitted value of $\beta = 0.65 \pm 0.15$ from the experimental data (solid green line).
The variation of $\bar{\rho}$ with $\varepsilon$ is compatible with that expected from DP in (2+1) dimensions, but the uncertainty in the imposed superheat prevents us from more accurately verifying the value of $\beta$.

\begin{figure}
  \centering
   \includegraphics[width=0.475\textwidth]{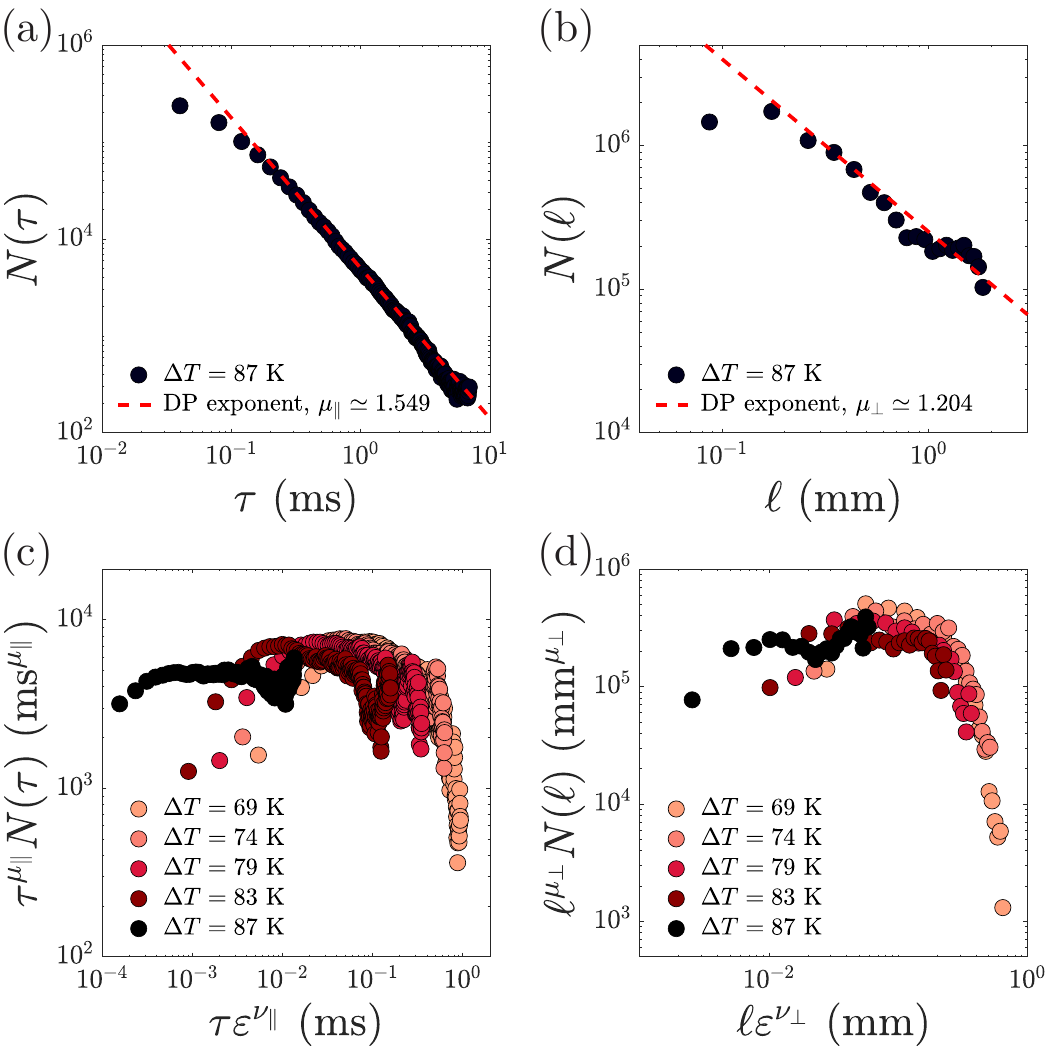}
   \caption{\label{fig4}
   (a-b) Distributions of the duration $\tau$ and lateral length $\ell$ of the dry intervals in the steady state for $\Delta T = 87$ K, close to the critical superheat. The distributions of $\ell$ and $\tau$ decay algebraically in good agreement with the DP power-law scalings (red dashed lines).
   \revP{(c-d) Collapse of the temporal (c) and spatial (d) distributions in a and b using the directed percolation power-laws: $\tau^{\mu^{\parallel}}N(\tau) \sim f(\tau\varepsilon^{\nu_{\parallel}})$ and $\ell^{\mu^{\perp}}N(\tau) \sim f(\ell\varepsilon^{\nu_{\perp}})$ where $f$ is an unknown function.}}
\end{figure}

To further assess whether the unpinning transition of the liquid/gas interface falls in the directed percolation universality class, we \revP{follow the approach of Takeuchi \emph{et al.} \cite{takeuchi2009}}, and explore the temporal and spatial distributions of the dry (inactive) regions close to criticality.
We define an inactive temporal interval $\tau$ (or similarly, an inactive spatial interval $\ell$) for any pixel $(i,j)$ in the observation area as an interval that verifies $h(i,j,t) < h_c$ and $h(i,j,t+\tau) < h_c$ while $h(i,j,t') > h_c$ for $0<t'<\tau$.
In figure \ref{fig4}a and b, we show the distributions $N(\tau)$ and $N(\ell)$ measured in the intermittent steady state for $\Delta T = 87$ K where $N(\tau)$ (respectively, $N(\ell)$) is the number of intervals of length $\tau$ (respectively, $\ell$).
Both distributions display power-law decays, in good agreement with what is expected from DP in (2+1) dimensions \cite{henkel2008}: namely, $N(\tau) \sim \tau^{-\mu_{\parallel}}$ and $N(\ell) \sim \ell^{-\mu_{\perp}}$ with $\mu_{\parallel} \simeq 1.549$ and $\mu_{\perp} \simeq 1.204$ \cite{grassberger1996} (dashed red lines).

Further away from the critical point, power-law scalings are expected only over a finite temporal and spatial range and the distributions have exponential tails (see figures 3a and b in the Supplemental Material).
For directed percolation, the correlation times, $\xi_\parallel$, and correlation lengths, $\xi_\perp$, associated with the distribution tails scale as $\xi_{\parallel,\perp} \sim \varepsilon^{-\nu_{\parallel,\perp}}$, where the exponents $\nu_{\parallel,\perp}$ and $\mu_{\parallel,\perp}$ are related by $\mu_{\parallel,\perp} = 2 - \beta/\nu_{\parallel,\perp}$ \cite{henkel2008}.
Merging the power-law and tail scalings, one obtains the generic form of the inactive interval distributions $N(\tau) \sim \tau^{-\mu_\parallel}f(\tau\varepsilon^{\nu_{\parallel}})$ and $N(\ell) \sim \ell^{-\mu_\perp}f(\ell\varepsilon^{\nu_{\perp}})$, where $f$ is an unknown function \cite{henkel2008}.
We test this scaling by plotting $\tau^{\mu_\parallel}N(\tau)$ and $\ell^{\mu_\perp}N(\ell)$ as a function of $\tau\varepsilon^{\nu_\parallel}$ and $\ell\varepsilon^{\nu_\perp}$ for superheats ranging from 69 K to 87 K taking the (2+1) dimensions DP exponents values $\nu_{\parallel} \simeq 1.295$ and $\nu_{\perp} \simeq 0.733$ \cite{grassberger1996} (\revP{figures \ref{fig4}c and d}).
The data collapses well for the spatial distributions but the collapse is less satisfying for the time intervals, a consequence of the limited observation time in our experiments that prevents us from observing exponential tails for all but the lowest superheat.

\begin{figure}
  \centering
   \includegraphics[width=0.475\textwidth]{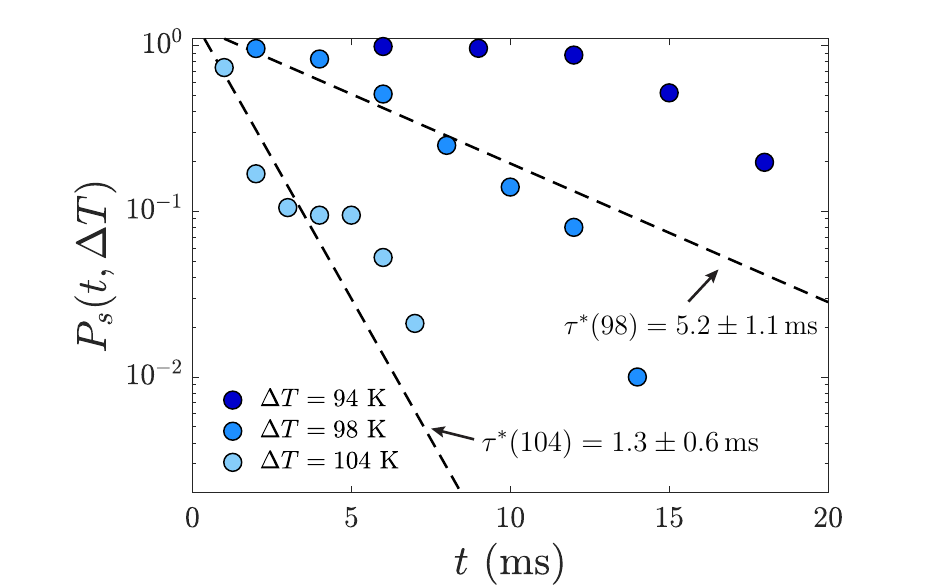}
   \caption{\label{fig5} \revP{Survival probability $P_s(t,\Delta T)$ of a contact nucleated at a single location during an impact for three different superheat $\Delta T > \Delta T_c$. The dashed lines are fits of the form $\mathrm{exp}[-(t-t_0)/\tau^*(\Delta T)]$, where $t_0$ is a constant and $\tau^*(\Delta T)$ is the characteristic lifetime of contact.}}
\end{figure}

Finally, we complement our steady state study by performing critical-spreading experiments, that is experiments in which wetting spreads from a single initial contact point.
We ensure that this condition is met by adjusting the impact velocity $U$, for each superheat, to be in the regime in which contamination can randomly puncture the vapor film and by discarding all impacts where multiple initial contact points are present (see Movie S2).
Such experiments allow us to measure the survival probability $P_s(t,\Delta T)$ that a cluster, which started from a single active seed, survives at time $t$.
We obtain statistics for the survival probability by realising $81-100$ impacts for each imposed value of the superheat $\Delta T$.
We immediately notice that the maximum observation time of approximately 18 ms, fixed by the drop rebound time, is too short to probe the power-law decay scaling of the survival probability at criticality.
We thus focus on the regime $\varepsilon < 0$, in which contact is short-lived, and show, in figure \ref{fig5}, the survival probability $P_s(t,\Delta T)$ for three superheats larger than $\Delta T_c$.
For $\Delta T = 98$ K and $104$ K, the survival times are indeed short enough to discard the influence of the finite observation time, and we observe that $P_s(t,\Delta T)$ decays approximately exponentially, a signature of a memoryless process \cite{barkley2016}.
The observed distribution can be described by a fit of the form $\mathrm{exp}[-(t-t_0)/\tau^*(\Delta T)]$ (black dashed lines) where $t_0$ is a constant related to the formation of the wetting pattern and $\tau^*(\Delta T)$ is the characteristic lifetime.
Clearly, $\tau^*(\Delta T)$ rapidly decreases as the superheat increases above the critical point, but our limited data does not allow to probe its functional form and possible divergence for a finite superheat \cite{hof2006,goldenfeld2010,barkley2016,kerswell2018}.
However, this qualitative picture, which holds for a wide range of impact velocities \cite{lee2020_2}, is key \revP{in the context of the dynamic Leidenfrost effect.
While contact can occur when $\Delta T > \Delta T_c$ for large impact velocities,}
only 10 K above the critical superheat, the characteristic lifetime of contact is much shorter than both the bouncing and evaporation time of the drop, indicating that heat transfer is largely reduced.

In this article, we have introduced an analogy between the transition towards the Leidenfrost state and phase transitions in the directed percolation universality class.
We reported the existence, in the presence of superheat, of the elementary mechanisms necessary for this analogy: the spreading and receding of transient wet patches coexisting with dry regions on the substrate in a spatiotemporal intermittency regime.
We then evidenced a continuous decrease of the steady state wet fraction with increasing superheat, leading to the unpinning of the liquid/gas interface from the substrate, and obtained statistics in agreement with the power-law scalings of the DP universality class.
This analogy allowed us to redefine the upper threshold of the Leidenfrost temperature $T_L^+$ as the critical point of a phase transition where two competing mechanisms (\emph{i.e}, the spreading and receding of wet regions) are balanced.
We stress that this definition of the Leidenfrost temperature does not stem from the hydrodynamic ability of the vapor layer to support the liquid but that it expresses the resilience of Leidenfrost drops to contact when $T>T_L^+$.
\revP{The insight gained from the identification of the DP behavior is not limited to the Leidenfrost temperature.
The absorbing nature of the levitated state indicates that, in the absence of perturbations, a floating drop will not spontaneously undergo a transition to the contact state, while the opposite is always allowed.
This asymmetry between contact and levitation is linked to the existence of metastable Leidenfrost drops down to the lower threshold of the Leidenfrost temperature $T_L^-$.}
\revP{Future experiments will hopefully be able to relieve the bouncing and cooling time constraints, allowing to further confirm the steady state approximation of this study, and to measure the survival probability close to the critical point.}
It would also be interesting to characterise the deterministic microscopic dynamics of the spreading and receding processes to build our understanding of the substrate and liquid properties that affect the balance between spreading and receding, and thus the value of the Leidenfrost temperature.

\begin{acknowledgments}
We acknowledge funding from the ERC Advanced Grant DDD under grant No. 740479.
\end{acknowledgments}

\bibliography{Bibli}

\end{document}